\documentclass[aps,preprint,showpacs,onecolumn,12pt]{revtex4}%
\usepackage{amsfonts}
\usepackage{amsmath}
\usepackage{amssymb}
\usepackage{graphicx}%
\begin{document}
\title{Alternative scheme for two-qubit conditional phase gate by adiabatic passage
under dissipation}
\author{Z. J. Deng$^{1,2}$}
\author{K. L. Gao$^{1}$}
\email{klgao@wipm.ac.cn}
\author{M. Feng$^{1}$}
\email{mangfeng1968@yahoo.com}
\affiliation{$^{1}$State Key Laboratory of Magnetic Resonance and Atomic and Molecular
Physics, Wuhan Institute of Physics and Mathematics, Chinese Academy of
Sciences, Wuhan 430071, China}
\affiliation{Centre for Cold Atom Physics, Chinese Academy of Sciences, Wuhan 430071, China}
\affiliation{$^{2}$Graduate School of the Chinese Academy of Sciences, Beijing 100049, China}

\begin{abstract}
We check a recent proposal [H. Goto and K. Ichimura Phys. Rev. A \textbf{70},
012305 (2004)] for controlled phase gate through adiabatic passage under the
influence of spontaneous emission and the cavity decay. We show a modification
of above proposal could be used to generate the necessary conditional phase
gates in the two-qubit Grover search. Conditioned on no photon leakage either
from the atomic excited state or from the cavity mode during the gating
period, we numerically analyze the success probability and the fidelity of the
two-qubit conditional phase gate by adiabatic passage. The comparison made
between our proposed gating scheme and a previous one shows that Goto and
Ichimura's scheme is an alternative and feasible way in the optical cavity
regime for two-qubit gates and could be generalised in principle to
multi-qubit gates.

\end{abstract}

\maketitle

\section{INTRODUCTION}

The stimulated Raman adiabatic passage (STIRAP) has been extensively studied
in coherent population transfer \cite{1}. By partially overlapping pulses in
the counterintuitive sequence, we have the population efficiently transferred
between two quantum states while almost not populating the intermediate level.
Thus the effect of spontaneous emission from the intermediate level is
negligible. Moreover, the STIRAP is independent of the pulse shape, which
makes it robust against moderate fluctuations of experimental parameters
\cite{2}. Because of the above mentioned merits, many schemes concerning
quantum information processing (QIP) are based on the technique of STIRAP,
such as single qubit rotation \cite{3}, controlled NOT gate \cite{4},
controlled phase gate \cite{5,6}, arbitrary state controlled-unitary gate
\cite{7}, SWAP gate \cite{8}, generation of qubit entanglement \cite{9} and so on.

All the above mentioned QIP schemes are quite different from the dynamical
ones \cite{10}, which need a precise control of the Rabi frequency and the
pulse duration. They are also quite different from the adiabatic geometric
ones \cite{11}, which depend on a controllable loop in the parameter space. In
Ref. \cite{5}, an attractive scheme to generate multi-qubit controlled phase
gate by only three steps in an optical cavity has been presented. As in Refs.
\cite{4,7,8}, it utilizes the cavity mode to couple different atomic qubits.
Although the dark states used for adiabatic evolution include no components of
atomic excited state, they usually contain one photon or multi-photon cavity
state. That is to say, the dark states are decoherence-free states with
respect to the spontaneous decay, but not immune to the cavity decay. However,
if the laser intensity is much smaller than the atom-cavity coupling strength
\cite{7,8}, the population of the cavity state with non-zero photon can be
very little. Nevertheless, with a weak laser intensity, the pulse duration
time must be much prolonged in order to fulfill the adiabatic condition
\cite{1,2}, which would enhance the probability of dissipation.

Therefore, it is natural for us to ask how well the quantum gating in Ref.
\cite{5} works under the influence of dissipation. In the present paper, we
will show how to generate all the conditional phase gates in a two-qubit
Grover search by the STIRAP technique, based on the main idea in Ref.
\cite{5}. To check the influence from spontaneous emission and the cavity
decay, we will employ the quantum jump approach \cite{12}, which uses wave
function to describe the system subject to dissipation while master equation
employs density matrix. Conditioned on no decay happening, the system's
evolution is governed by a non-Hermitian Hamiltonian. In some cases, quantum
jump approach could help us find analytical solutions for systems subject to
dissipation \cite{13}. But in the present case, due to complexity, we will
have to numerically simulate the gating process and investigate the dependence
of the success probability and the gate fidelity on the spontaneous emission
rate $\Gamma$ and the cavity decay rate $\kappa$.

\section{CONDITIONAL PHASE GATE WITHOUT DECAY}

The conditional phase gate is used for labeling target state in the Grover
search algorithm \cite{14,15}, which means an addition of a $\pi$ phase as a
prefactor to the target state, but of nothing to other states. The target
state labeling is a key step in Grover search. For items represented by the
computational states $|X\rangle$ with $X=0,1,...,N-1,$ in a quantum register
with $n$ qubits, we have $N=2^{n}$ possible states. If the target state is
$|\tau\rangle$, the conditional phase gate can be expressed as $I_{\tau}=$
$I-2|\tau\rangle\langle\tau|$, where $I$ is the $N\times N$ identity matrix.
By redefining the energy levels in Ref. \cite{5}, we show that the second step
in Ref. \cite{5} to generate the two-qubit controlled phase gate is actually
for a conditional phase gating to label the target state $|0\rangle
_{1}|1\rangle_{2}$. Atoms are fixed in the optical cavity, as shown in Fig. 1,
for an atom j with five-level configuration, where levels $|0\rangle_{j}$ and
$|\sigma\rangle_{j}$ are coupled to the excited state $|2\rangle_{j}$ by two
lasers with Rabi frequencies $\Omega_{0,j}$ and $\Omega_{\sigma,j}$
respectively, while the level $|1\rangle_{j}$ is coupled to state
$|2\rangle_{j}$ by the cavity mode with the coupling constant $g_{j}$. These
three couplings are needed to construct a two-qubit or even a multi-qubit
gate. The couplings of the levels $|0\rangle_{j}$, $|\sigma\rangle_{j}$ and
$|1\rangle_{j}$ to another excited state $|3\rangle_{j}$ can be used to
perform single qubit rotation as in Ref. \cite{3}.

We encode the qubits in levels $|0\rangle_{j}$ and $|1\rangle_{j}$, and for
simplicity we focus our discussion on the case of two atoms, although our case
is extendable to multi-atom cases. So our task is to accomplish $I_{|i\rangle
_{1}|j\rangle_{2}},$ i.e., adding a minus sign to the target states
$|i\rangle_{1}|j\rangle_{2}$ where i, j being 0, 1, respectively and the
subscripts are for different atoms. Consider the pulses regarding
$\Omega_{0,1}$ and $\Omega_{\sigma,2}$, the Hamiltonian is given by (assuming
$\hbar=1$)%

\begin{equation}
H=\Omega_{0,1}|2\rangle_{11}\langle0|+\Omega_{\sigma,2}|2\rangle_{22}%
\langle\sigma|+\underset{m=1,2}{\sum}g_{m}a|2\rangle_{mm}\langle1|+H.c.,
\label{1}%
\end{equation}
where $a$ is the annihilation operator for the cavity mode. As the cavity is
initially in the vacuum state $|0\rangle$, $|1\rangle_{1}|0\rangle
_{2}|0\rangle$ and $|1\rangle_{1}|1\rangle_{2}|0\rangle$ are unaffected by the
Hamiltonian, while $|0\rangle_{1}|0\rangle_{2}|0\rangle$ and $|0\rangle
_{1}|1\rangle_{2}|0\rangle$ are associated with the Hamiltonian's two dark
states (i.e., eigenstates with zero eigenvalues) $|D_{00}\rangle\propto
g_{1}|0\rangle_{1}|0\rangle_{2}|0\rangle-\Omega_{0,1}|1\rangle_{1}%
|0\rangle_{2}|1\rangle$, $|D_{01}\rangle\propto g_{1}\Omega_{\sigma
,2}|0\rangle_{1}|1\rangle_{2}|0\rangle+g_{2}\Omega_{0,1}|1\rangle_{1}%
|\sigma\rangle_{2}|0\rangle-\Omega_{0,1}\Omega_{\sigma,2}|1\rangle
_{1}|1\rangle_{2}|1\rangle$ respectively. The procedure consists of two STIRAP
processes: (i) In the first STIRAP, $\Omega_{\sigma,2}$ precedes $\Omega
_{0,1}$; (ii) The second STIRAP is a reverse process of the first one but with
the phase regarding $\Omega_{\sigma,2}$ added by $\pi$. In the adiabatic limit
and on the condition that the Berry phase is equal to zero, we have \cite{5}
(assuming $\Omega_{0,1}$, $\Omega_{\sigma,2}$ having the same phase in (i) and
$g_{1}=g_{2}=g$)
\begin{align}
&  |0\rangle_{1}|0\rangle_{2}|0\rangle\overset{\text{(i)}}{\rightarrow
}|0\rangle_{1}|0\rangle_{2}|0\rangle\overset{\text{(ii)}}{\rightarrow
}|0\rangle_{1}|0\rangle_{2}|0\rangle\label{2}\\
&  |0\rangle_{1}|1\rangle_{2}|0\rangle\overset{\text{(i)}}{\rightarrow
}|1\rangle_{1}|\sigma\rangle_{2}|0\rangle\overset{\text{(ii)}}{\rightarrow
}-|0\rangle_{1}|1\rangle_{2}|0\rangle\nonumber\\
&  |1\rangle_{1}|0\rangle_{2}|0\rangle\overset{\text{(i)}}{\rightarrow
}|1\rangle_{1}|0\rangle_{2}|0\rangle\overset{\text{(ii)}}{\rightarrow
}|1\rangle_{1}|0\rangle_{2}|0\rangle\nonumber\\
&  |1\rangle_{1}|1\rangle_{2}|0\rangle\overset{\text{(i)}}{\rightarrow
}|1\rangle_{1}|1\rangle_{2}|0\rangle\overset{\text{(ii)}}{\rightarrow
}|1\rangle_{1}|1\rangle_{2}|0\rangle\nonumber
\end{align}
Thus the conditional phase gate for labeling the target state $|0\rangle
_{1}|1\rangle_{2}$ is generated. All the other conditional phase gates can be
generated by adding the NOT gate on both sides of the above two STIRAP
processes. It is easy to see $I_{|0\rangle_{1}|0\rangle_{2}}=\sigma
_{x,2}I_{|0\rangle_{1}|1\rangle_{2}}\sigma_{x,2}$, $I_{|1\rangle_{1}%
|0\rangle_{2}}=\sigma_{x,1}\sigma_{x,2}I_{|0\rangle_{1}|1\rangle_{2}}%
\sigma_{x,2}\sigma_{x,1}$, $I_{|1\rangle_{1}|1\rangle_{2}}=\sigma
_{x,1}I_{|0\rangle_{1}|1\rangle_{2}}\sigma_{x,1}$, where $\sigma_{x,i}$
$(i=1,2)$ is the NOT gate acting on the $i$th atom, transforming states as
$|0\rangle_{i}\rightarrow|1\rangle_{i}$, $|1\rangle_{i}\rightarrow
|0\rangle_{i}$. The NOT gate $\sigma_{x,i}$ can be obtained by coupling
$|0\rangle_{i}$, $|\sigma\rangle_{i}$ and $|1\rangle_{i}$ with the excited
state $|3\rangle_{i}$, by three STIRAP processes: (1) $\overset{\sim}{\Omega
}_{\sigma,i}$ precedes $\overset{\sim}{\Omega}_{1,i}$; (2) $\overset{\sim
}{\Omega}_{1,i}$ precedes $\overset{\sim}{\Omega}_{0,i}$; (3) $\overset{\sim
}{\Omega}_{0,i}$ precedes $\overset{\sim}{\Omega}_{\sigma,i}$. The dark states
associated with the above three steps are $|D^{\text{(1)}}\rangle
\propto\overset{\sim}{\Omega}_{\sigma,i}|1\rangle_{i}-\overset{\sim}{\Omega
}_{1,i}|\sigma\rangle_{i}$, $|D^{\text{(2)}}\rangle\propto\overset{\sim
}{\Omega}_{1,i}|0\rangle_{i}-\overset{\sim}{\Omega}_{0,i}|1\rangle_{i}$,
$|D^{\text{(3)}}\rangle\propto\overset{\sim}{\Omega}_{0,i}|\sigma\rangle
_{i}-\overset{\sim}{\Omega}_{\sigma,i}|0\rangle_{i}$ respectively. In the
adiabatic limit and with the condition of zero Berry phase, we obtain in the
case of the two pulses in each STIRAP with $\pi$ phase difference,%

\begin{align}
&  |0\rangle_{i}\overset{\text{(1)}}{\rightarrow}|0\rangle_{i}\overset
{\text{(2)}}{\rightarrow}|1\rangle_{i}\overset{\text{(3)}}{\rightarrow
}|1\rangle_{i},\label{3}\\
&  |1\rangle_{i}\overset{\text{(1)}}{\rightarrow}|\sigma\rangle_{i}%
\overset{\text{(2)}}{\rightarrow}|\sigma\rangle_{i}\overset{\text{(3)}%
}{\rightarrow}|0\rangle_{i}\text{.}\nonumber
\end{align}
As a result, we realize the NOT gate. Alternatively, the NOT gate can also be
reached by single qubit rotation as in Ref. \cite{3}, which also needs 6 pulses.

\section{CONDITIONAL PHASE GATE WITH DECAY}

We introduce the spontaneous emission and the cavity decay from now on and
will numerically analyze their effects on the conditional phase gate by the
quantum jump approach. As long as there is no photon leakage either from the
atomic excited state or from the cavity mode during the gating period, the
Hamiltonian in Eq. (1) becomes%

\begin{equation}
H_{cond}=[\Omega_{0,1}|2\rangle_{11}\langle0|+\Omega_{\sigma,2}|2\rangle
_{22}\langle\sigma|+\underset{m=1,2}{\sum}ga|2\rangle_{mm}\langle
1|+H.c.]-i\frac{\kappa}{2}a^{\dagger}a-i\frac{\Gamma}{2}\underset{m=1,2}{\sum
}|2\rangle_{mm}\langle2|, \label{4}%
\end{equation}
where we have made $g_{1}=g_{2}=g$. The above Hamiltonian can be written as a
matrix in the subspace spanned by $|0\rangle_{1}|1\rangle_{2}|0\rangle$,
$|2\rangle_{1}|1\rangle_{2}|0\rangle$, $|1\rangle_{1}|1\rangle_{2}|1\rangle$,
$|1\rangle_{1}|2\rangle_{2}|0\rangle$, $|1\rangle_{1}|\sigma\rangle
_{2}|0\rangle$, $|0\rangle_{1}|0\rangle_{2}|0\rangle$, $|2\rangle_{1}%
|0\rangle_{2}|0\rangle$, $|1\rangle_{1}|0\rangle_{2}|1\rangle$, $|1\rangle
_{1}|0\rangle_{2}|0\rangle$, $|1\rangle_{1}|1\rangle_{2}|0\rangle$%

\begin{equation}
H_{cond}=\left[
\begin{array}
[c]{cccccccccc}%
0 & \Omega_{0,1}^{\ast} & 0 & 0 & 0 & 0 & 0 & 0 & 0 & 0\\
\Omega_{0,1} & -i\frac{\Gamma}{2} & g & 0 & 0 & 0 & 0 & 0 & 0 & 0\\
0 & g^{\ast} & -i\frac{\kappa}{2} & g^{\ast} & 0 & 0 & 0 & 0 & 0 & 0\\
0 & 0 & g & -i\frac{\Gamma}{2} & \Omega_{\sigma,2} & 0 & 0 & 0 & 0 & 0\\
0 & 0 & 0 & \Omega_{\sigma,2}^{\ast} & 0 & 0 & 0 & 0 & 0 & 0\\
0 & 0 & 0 & 0 & 0 & 0 & \Omega_{0,1}^{\ast} & 0 & 0 & 0\\
0 & 0 & 0 & 0 & 0 & \Omega_{0,1} & -i\frac{\Gamma}{2} & g & 0 & 0\\
0 & 0 & 0 & 0 & 0 & 0 & g^{\ast} & -i\frac{\kappa}{2} & 0 & 0\\
0 & 0 & 0 & 0 & 0 & 0 & 0 & 0 & 0 & 0\\
0 & 0 & 0 & 0 & 0 & 0 & 0 & 0 & 0 & 0
\end{array}
\right]  . \label{5}%
\end{equation}
Note that $H_{cond}$ is non-Hermitian, the norm of a state vector evolving
under the corresponding Schr\"{o}dinger equation decreases in general with
time. For an arbitrary initial state $|\psi(t_{i})\rangle$,
\begin{equation}
P_{suc}(t)=\langle\psi(t_{i})|U_{cond}^{\dagger}(t,t_{i})U_{cond}%
(t,t_{i})|\psi(t_{i})\rangle\label{6}%
\end{equation}
defines the probability that no photon has been emitted at time $t$, where
$U_{cond}$ is the time evolution operator for $H_{cond}$ and the corresponding
normalized state vector is%

\begin{equation}
|\psi(t)\rangle=U_{cond}(t,t_{i})|\psi(t_{i})\rangle/\sqrt{P_{suc}(t)}.
\label{7}%
\end{equation}
The gate fidelity is%

\begin{equation}
F=|\langle\psi(\infty)|I_{|0\rangle_{1}|1\rangle_{2}}|\psi(t_{i})\rangle|^{2}.
\label{8}%
\end{equation}

In our numerical simulation, we choose $\Omega_{0,1,\max}=\Omega
_{\sigma,2,\max}=0.16g$ in order to reduce the population of single-photon
cavity states during the gating process, i.e., $|1\rangle_{1}|0\rangle
_{2}|1\rangle$ and $|1\rangle_{1}|1\rangle_{2}|1\rangle$. Similar to Refs.
\cite{3,5}, we suppose that the pulses have the Gaussian shape $\exp[-(t\pm
T/2\pm t_{0})^{2}/2\tau^{2}]$ with $T=200/g$, $t_{0}=30/g$, and $\tau=40/g$.
In Fig. 2, we show the time evolution of the initial state $\frac{1}%
{2}(|0\rangle_{1}|0\rangle_{2}+|0\rangle_{1}|1\rangle_{2}+|1\rangle
_{1}|0\rangle_{2}+|1\rangle_{1}|1\rangle_{2})|0\rangle$. Ideally, the final
state would be $\frac{1}{2}(|0\rangle_{1}|0\rangle_{2}-|0\rangle_{1}%
|1\rangle_{2}+|1\rangle_{1}|0\rangle_{2}+|1\rangle_{1}|1\rangle_{2})|0\rangle
$. By our numerical calculation, with $k=0.1g$ and $\Gamma=0.1g$, the
probability amplitudes for states $|0\rangle_{1}|0\rangle_{2}|0\rangle$,
$|0\rangle_{1}|1\rangle_{2}|0\rangle$, $|1\rangle_{1}|0\rangle_{2}|0\rangle$,
$|1\rangle_{1}|1\rangle_{2}|0\rangle$ are 0.4513, -0.4523, 0.5438, 0.5438
respectively, and the fidelity is 99.12\% while the success probability is
84.55\%. The reason for the amplitudes of the components $|1\rangle
_{1}|0\rangle_{2}|0\rangle$ and $|1\rangle_{1}|1\rangle_{2}|0\rangle$
increased compared with those in the initial state is that they did not
participate in the evolution, while the other two components dissipated in the
evolution. So the relative weights of $|1\rangle_{1}|0\rangle_{2}|0\rangle$
and $|1\rangle_{1}|1\rangle_{2}|0\rangle$ are enlarged in the normalized final state.

In order to see how the success probability $P_{suc}$ and the gate fidelity
$F$ depend on $\kappa$ and $\Gamma$, we plot $P_{suc}$ and $F$ in each of
Figs. 3, 4, 5, for the evolution from the initial states $|0\rangle
_{1}|1\rangle_{2}|0\rangle$, $|0\rangle_{1}|0\rangle_{2}|0\rangle$, $\frac
{1}{2}(|0\rangle_{1}|0\rangle_{2}+|0\rangle_{1}|1\rangle_{2}+|1\rangle
_{1}|0\rangle_{2}+|1\rangle_{1}|1\rangle_{2})|0\rangle$, respectively. From
Fig. 3, we see that $P_{suc}$ is less affected by $\kappa$ than by $\Gamma$,
while it is almost unaffected by $\Gamma$ in Fig. 4. Since for the evolution
of $|0\rangle_{1}|1\rangle_{2}|0\rangle,$ the occupation probability of
$|1\rangle_{1}|1\rangle_{2}|1\rangle$ is several times smaller than that of
the excited atomic state, the spontaneous emission from the excited state
$|2\rangle$ is more detrimental than cavity decay in Fig. 3. While for the
evolution of $|0\rangle_{1}|0\rangle_{2}|0\rangle$ in Fig. 4, as the
occupation in level
$\vert$%
$2\rangle_{1}$ is negligible, the cavity decay is the main detrimental effect
in this case. For an arbitrary initial state, $|\psi(t_{i})\rangle
=\alpha|0\rangle_{1}|1\rangle_{2}|0\rangle+\beta|0\rangle_{1}|0\rangle
_{2}|0\rangle+\gamma|1\rangle_{1}|0\rangle_{2}|0\rangle+\varepsilon
|1\rangle_{1}|1\rangle_{2}|0\rangle$, we have $U_{cond}(\infty,t_{i}%
)|\psi(t_{i})\rangle=$ $\alpha\sqrt{P_{suc1}}|\psi_{010}\rangle+\beta
\sqrt{P_{suc2}}|\psi_{000}\rangle+\gamma|1\rangle_{1}|0\rangle_{2}%
|0\rangle+\varepsilon|1\rangle_{1}|1\rangle_{2}|0\rangle$, where $|\psi
_{010}\rangle$ and $|\psi_{000}\rangle$ are, respectively, the final states
after time evolution of the component states $|0\rangle_{1}|1\rangle
_{2}|0\rangle$, $|0\rangle_{1}|0\rangle_{2}|0\rangle,$ and $P_{suc1}$,
$P_{suc2}$ are their corresponding success probabilities. Thus the success
probability for any initial state is $P_{suc}=|\alpha|^{2}P_{suc1}+|\beta
|^{2}P_{suc2}+|\gamma|^{2}+|\varepsilon|^{2}$. Usually both $P_{suc1}$ and
$P_{suc2}$ are smaller than $1$. As a result, for the same $\Gamma$ and
$\kappa$, the less the initial population in components $|0\rangle
_{1}|1\rangle_{2}|0\rangle$ and $|0\rangle_{1}|0\rangle_{2}|0\rangle$, the
higher the success probability for the gating process. Moreover, from Figs. 3
and 4, we know that the fidelity is very high for the initial state
$|0\rangle_{1}|1\rangle_{2}|0\rangle$ or $|0\rangle_{1}|0\rangle_{2}|0\rangle
$, which implies $|\psi_{010}\rangle\simeq-|0\rangle_{1}|1\rangle_{2}%
|0\rangle$, $|\psi_{000}\rangle\simeq|0\rangle_{1}|0\rangle_{2}|0\rangle$,
while the relative weight ratio of the four component states $|0\rangle
_{1}|1\rangle_{2}|0\rangle$, $|0\rangle_{1}|0\rangle_{2}|0\rangle$,
$|1\rangle_{1}|0\rangle_{2}|0\rangle$, $|1\rangle_{1}|1\rangle_{2}|0\rangle$
is approximately $(-\alpha\sqrt{P_{suc1}}$, $\beta\sqrt{P_{suc2}}$, $\gamma$,
$\varepsilon)$, deviated from the ideal case $(-\alpha$, $\beta$, $\gamma$,
$\varepsilon)$. Obviously, the fidelity for the state $|\psi(t_{i})\rangle$
would be very high in the following two cases: 1) $\gamma=\varepsilon=0$,
$P_{suc1}\simeq P_{suc2}$; 2) $\gamma$, $\varepsilon\neq0$, $P_{suc1}\simeq
P_{suc2}\simeq1$. Therefore, it is easy to understand in Fig. 5 why $P_{suc}$
is increased in most regions while $F$ is decreased by comparing with those in
Figs. 3 and 4 with the same $\Gamma$ and $\kappa$.

We also did the simulation for the NOT gate (not shown), which shows that with
intense short pulses and the vacuum cavity state, the fidelity and the success
probability are both near unity even under large $\kappa$ and $\Gamma$. So the
success probability and\ the fidelity for the other three conditional phase
gates would have almost the same dependence on $\kappa$ and $\Gamma$ as in
above figures.

\section{DISCUSSION AND CONCLUSION}

Combined with single qubit Hadamard gate, which is achievable by a series of
single qubit rotations as in Ref. \cite{3}, the performance of the two-qubit
Grover search by the conditional phase gates is straightforward \cite{14,15}.
Our scheme has the following advantages compared with Ref. \cite{15}: Atoms
are fixed in the cavity, which is easier to control than the case with atoms
going through the cavity with a certain velocity. Moreover, all the operations
are based on STIRAP, so they are very robust to fluctuation of experimental parameters.

We now briefly discuss the experimental feasibility. It has been
experimentally reported that $g=34\times2\pi$ MHz, $\kappa=4.1\times2\pi$ MHz
and $\Gamma=2.6\times2\pi$ MHz \cite{16}. If we adopt the above numbers, i.e.,
$\kappa/g\simeq0.12$, $\Gamma/g\simeq0.077$, the operation time is of the
order of 10$^{-6}$ s, as can be seen from Fig. 2(a). In this case, for the
initial states $|0\rangle_{1}|1\rangle_{2}|0\rangle$, $|0\rangle_{1}%
|0\rangle_{2}|0\rangle$ and $\frac{1}{2}(|0\rangle_{1}|0\rangle_{2}%
+|0\rangle_{1}|1\rangle_{2}+|1\rangle_{1}|0\rangle_{2}+|1\rangle_{1}%
|1\rangle_{2})|0\rangle$, we may obtain that $P_{suc}\simeq73.1\%$, $63.8\%$,
$84.2\%$, and $F\simeq99.9\%$, $100\%$, $99.0\%$, respectively. The fidelities
are high enough, while the success probabilities need to be improved by
reducing $\kappa/g$ and $\Gamma/g$ in the future experiment.

We have noticed a previous smart scheme \cite{17} for two-qubit phase gate,
based on the STIRAP technique, by an almost same configuration as in the
present paper. Assisted by Zeno effect, the state involved in adiabatic
evolution in Ref. \cite{17} is immune to both the cavity decay and spontaneous
emission. From their simulation for a STIRAP process with fixed parameters
$k=0.1g$ and $\Gamma=0.1g$, we see the variation of the maximum Rabi frequency
and the detuning of the corresponding transitions of the atomic levels can
lead to the maximum success probability and the high fidelity for the
population transfer \cite{18}. In contrast, we did simulations in our work for
successive two STIRAP processes without any change of the detuning and the
pulse Rabi frequency \cite{19}. Given all the facts above and by comparing the
simulation results, we consider that the proposal presented in Ref. \cite{5}
is not less effective than that in Ref. \cite{17} in the presence of
dissipation. Moreover, Ref. \cite{5} is applicable to multi-qubit phase gates,
which would be used in multi-qubit Grover search algorithm. But with more
qubits involved, the dark states containing components with more photons,
would be more sensitive to dissipation. However, with very high Q cavity, Ref.
\cite{5} can be in principle feasible in the case of multi-qubit phase gates.

In summary, we have shown a generation of four conditional phase gates
required by a two-qubit Grover search algorithm by means of the STIRAP
technique, in the presence of dissipation. To meet the requirement of
adiabatic condition, the multiplication of the Rabi frequency and the
implementation time should be much larger than unity. But the larger the Rabi
frequency, the more the occupation probability in the cavity mode, yielding
more probability of the cavity decay. On the other hand, the less Rabi
frequency would require a longer implementational time, which would also
enhance the probability of dissipation. In this sense, it is of importance to
find suitable numbers of above parameters. The numbers are fixed in our
simulation, it would not be optimal for all the combinative values of $k/g$
and $\Gamma/g$. But we have tried to used the numbers, which could result in
relatively large values of $P_{suc}$ and $F$ in the whole range of $k/g$ and
$\Gamma/g$. We believe that our numerical investigation for the success
probability and the fidelity of the two-qubit conditional phase gate subject
to the spontaneous emission and the cavity decay would be useful for real experiments.

{\large ACKNOWLEDGMENTS}

Z. J. Deng is grateful to Weibin Li, T. Y. Shi and X. L. Zhang for their
warmhearted help. This work is partly supported by National Natural Science
Foundation of China under Grant Nos. 10474118 and 60490280, by Hubei
Provincial Funding for Distinguished Young Scholar, and partly by the National
Fundamental Research Program of China under Grant No. 2005CB724502.


\begin{thebibliography}{99}
\bibitem {18}The simultion in Ref. \cite{17} only involved one STIRAP process
for population transfer. As a result, the optimal success probability and
fidelity in the population transfer with fixed parameters $k=0.1g$ and
$\Gamma=0.1g$ are 0.858, 0.999 respectively. For a complete controlled phase
gate, however, there should be two successive STIRAP processes.
\bibitem {19}For given values of $k/g$ and $\Gamma/g$, we may vary
$\Omega_{0,1,\max}/g$, $\Omega_{\sigma,2,\max}/g$, $gt_{0}$, $g\tau$, the
transition detunings to get the optimal values of $P_{suc}$ and $F$. However,
our purpose is to illustrate how $P_{suc}$ and $F$ depend on $k/g$ and
$\Gamma/g$. It's troublesome to adjust all the above mentioned parameters to
find the optimal values of $P_{suc}$ and $F$ for each combinative values of
$k/g$ and $\Gamma/g$. On the contrary, with fixed parameters, we can analyze
the underlying physics of why $P_{suc}$ and $F$ vary with $k/g$ and $\Gamma/g$
as plotted in Figs. 3, 4, 5. So we fix the detunings to be zero and all the
parameters for the pulse configuration to be the ones as plotted in Fig. 2(a).

\newpage

{\LARGE Figure captions}

FIG. 1. The level configuration for atom j, where the qubits are encoded in
levels $|0\rangle_{j}$ and $|1\rangle_{j}$. Except that levels $|1\rangle_{j}$
and $|2\rangle_{j}$ are coupled resonantly by the cavity mode, all the other
arrows indicate resonant couplings by different lasers with the corresponding
Rabi frequencies labeled nearby.

FIG. 2. Time evolution with the initial state $\frac{1}{2}(|0\rangle
_{1}|0\rangle_{2}+|0\rangle_{1}|1\rangle_{2}+|1\rangle_{1}|0\rangle
_{2}+|1\rangle_{1}|1\rangle_{2})|0\rangle$, where $k=0.1g$ and $\Gamma=0.1g$.
(a) Pulse shapes and sequences for $\Omega_{0,1}$ and $\Omega_{\sigma,2}$. (b)
Probability amplitudes for states $|0\rangle_{1}|1\rangle_{2}|0\rangle$,
$|1\rangle_{1}|\sigma\rangle_{2}|0\rangle$, $|0\rangle_{1}|0\rangle
_{2}|0\rangle$, $|1\rangle_{1}|0\rangle_{2}|0\rangle$, $|1\rangle_{1}%
|1\rangle_{2}|0\rangle$. (c) Probability of all the states containing excited
atomic state, i.e., $P_{exc}=P_{|2\rangle_{1}|1\rangle_{2}|0\rangle
}+P_{|1\rangle_{1}|2\rangle_{2}|0\rangle}+P_{|2\rangle_{1}|0\rangle
_{2}|0\rangle}$ , and probability of all the states containing one photon,
i.e., $P_{pht}=P_{|1\rangle_{1}|1\rangle_{2}|1\rangle}+P_{|1\rangle
_{1}|0\rangle_{2}|1\rangle}$. (d) Success probability $P_{suc}$.

FIG. 3. Dependence of success probability $P_{suc}$ (upper subplot) and
fidelity $F$ (lower subplot) on $\kappa$ and $\Gamma$ for the initial state
$|0\rangle_{1}|1\rangle_{2}|0\rangle$.

FIG. 4. Dependence of success probability $P_{suc}$ (upper subplot) and
fidelity $F$ (lower subplot) on $\kappa$ and $\Gamma$ for the initial state
$|0\rangle_{1}|0\rangle_{2}|0\rangle$.

FIG. 5. Dependence of success probability $P_{suc}$ (upper subplot) and
fidelity $F$ (lower subplot) on $\kappa$ and $\Gamma$ for the initial state
$\frac{1}{2}(|0\rangle_{1}|0\rangle_{2}+|0\rangle_{1}|1\rangle_{2}%
+|1\rangle_{1}|0\rangle_{2}+|1\rangle_{1}|1\rangle_{2})|0\rangle$.
\end{thebibliography}
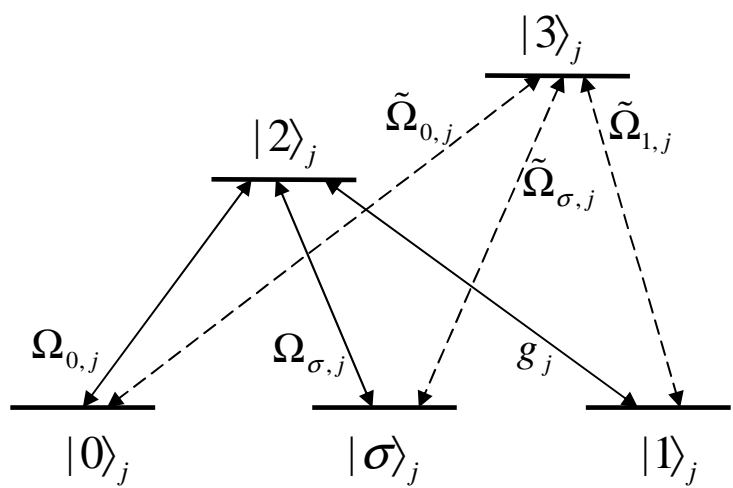
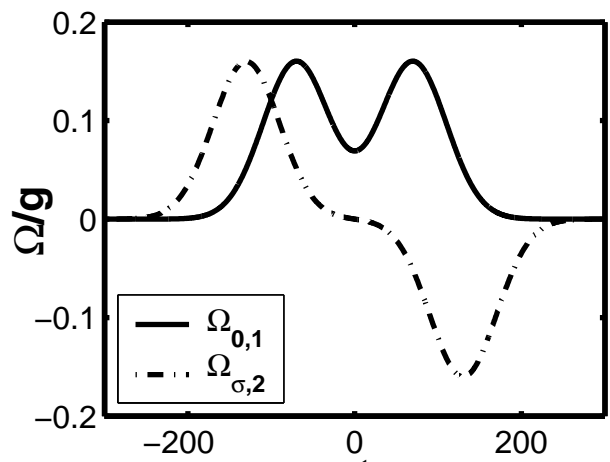
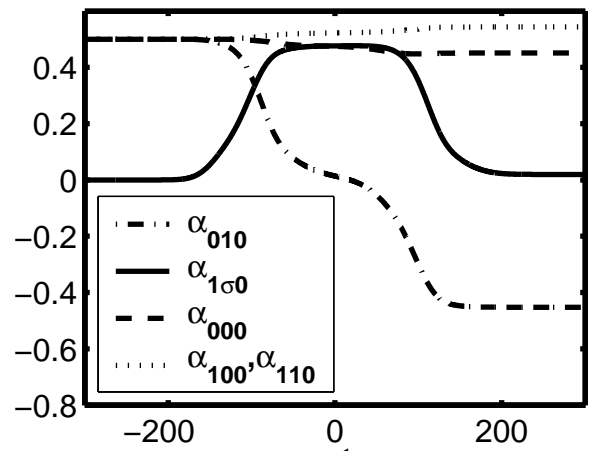
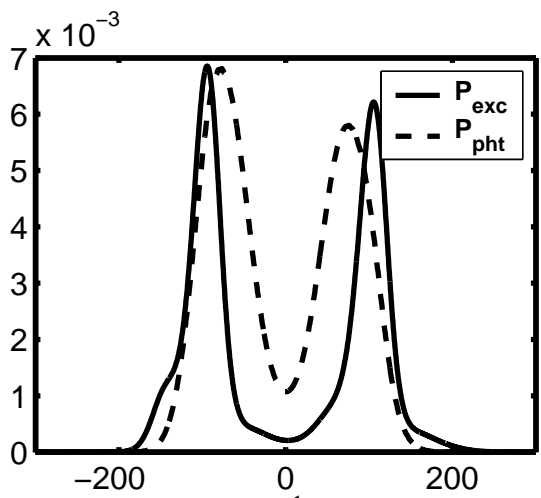
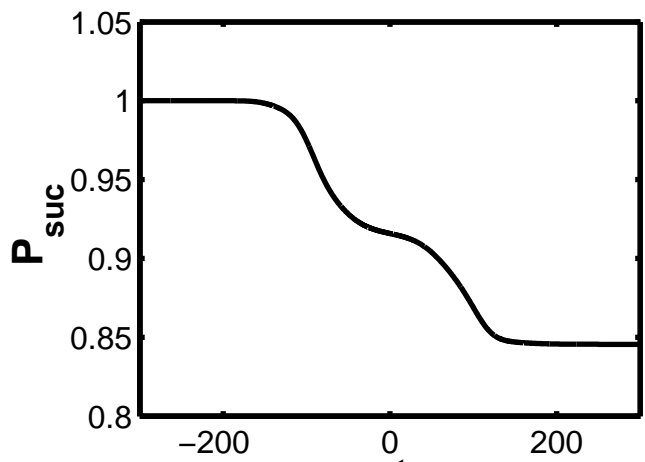
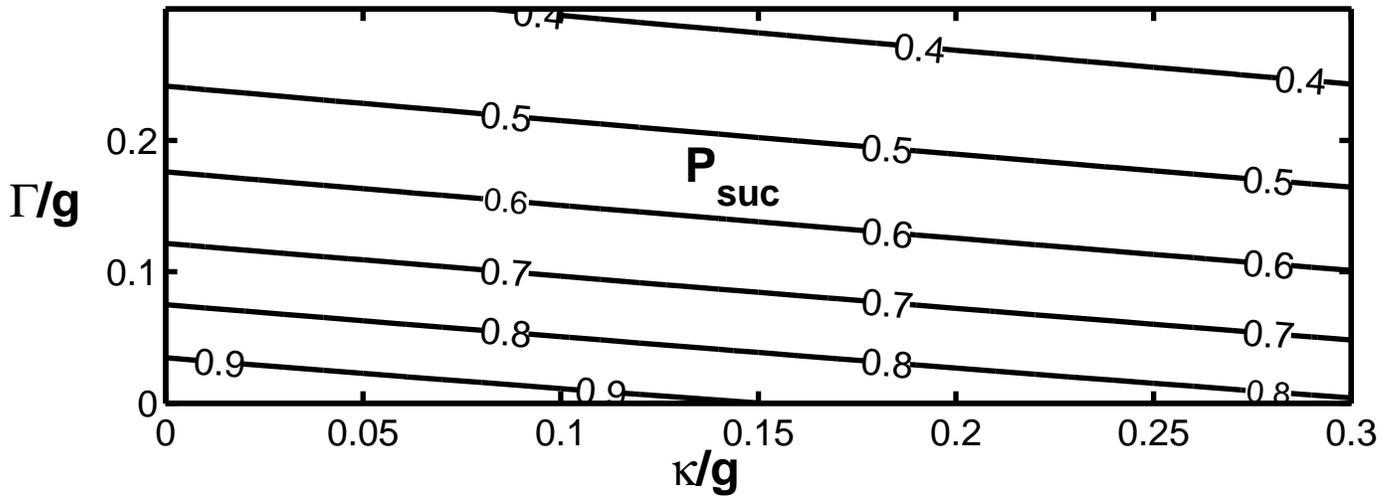
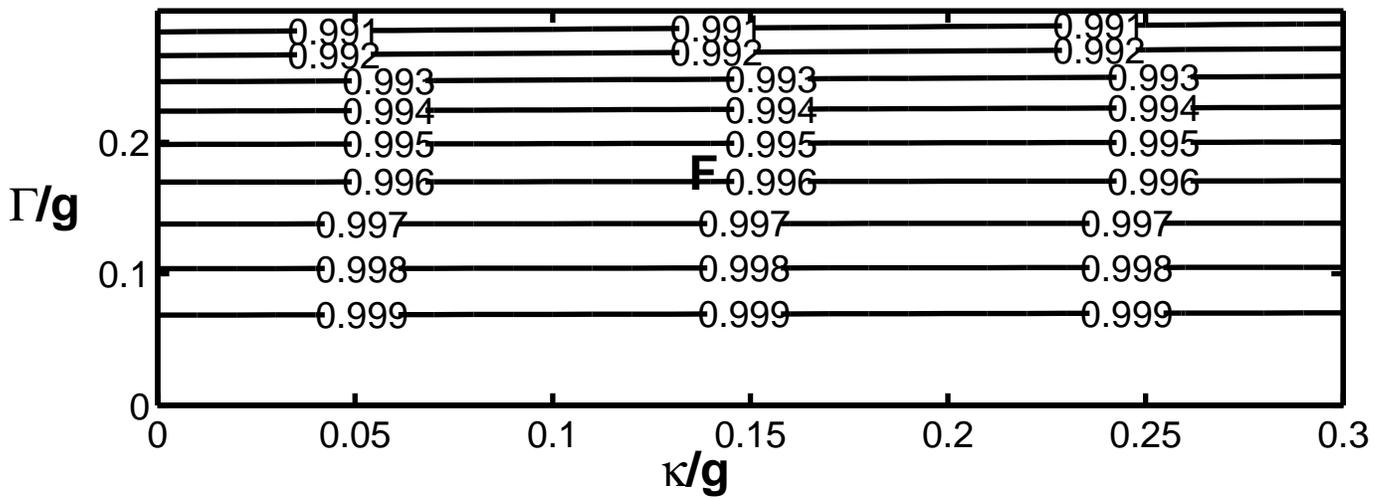
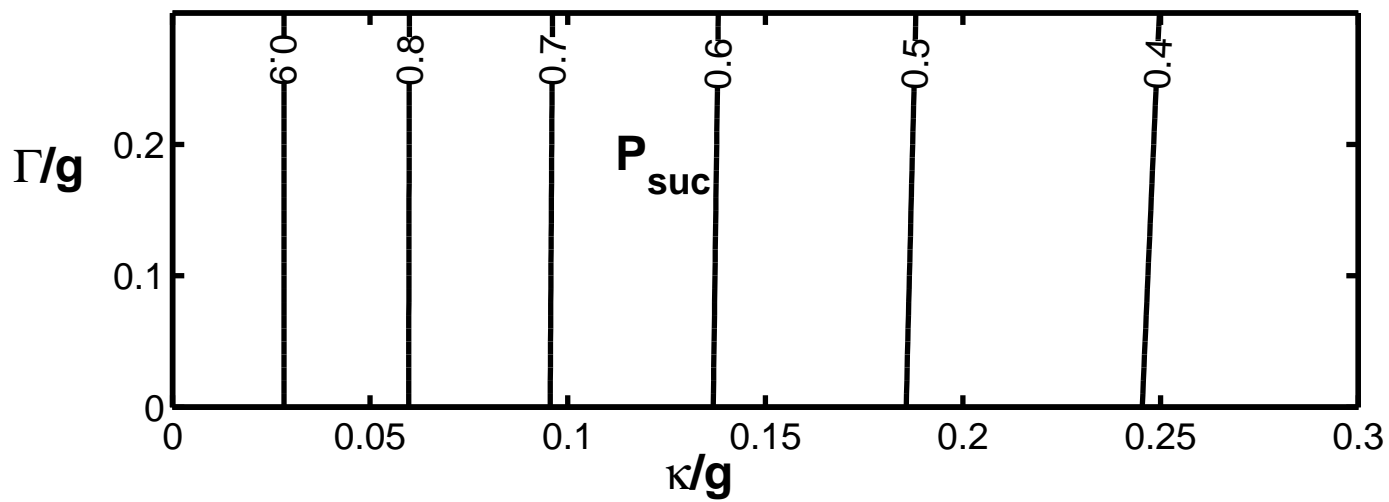
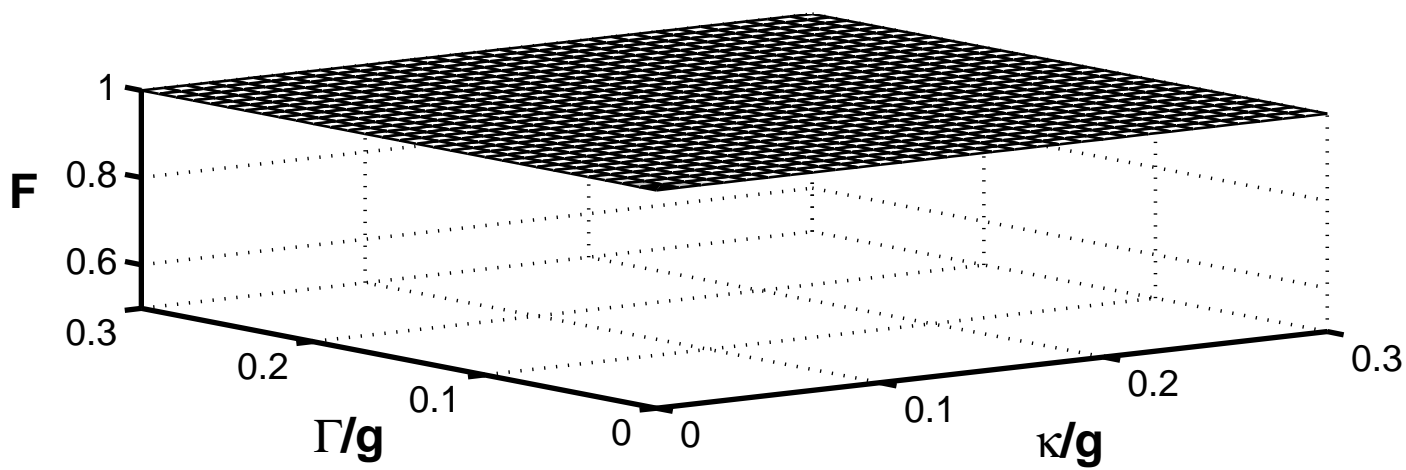
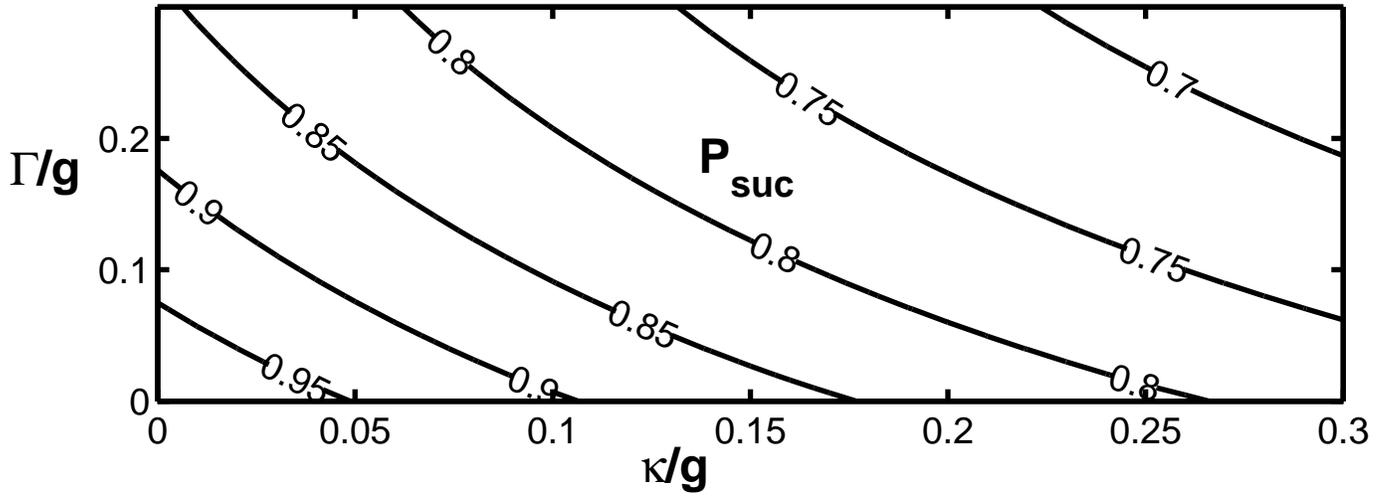
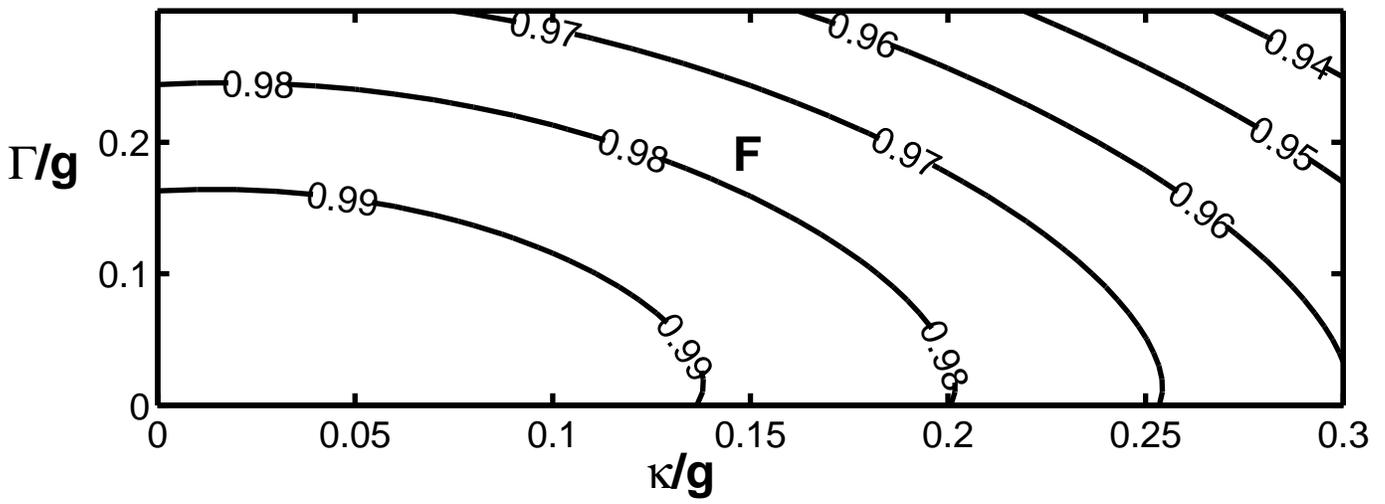
\end{document}